\documentclass{article}
\usepackage{spconf,amsmath,graphicx}

\title{DENOISING CONVOLUTIONAL AUTOENCODER BASED B-MODE ULTRASOUND TONGUE IMAGE FEATURE EXTRACTION}

\name{Bo Li$^{1}$,
	Kele Xu$^{2,3}$\sthanks{Corresponding author.},
	Dawei Feng$^{2,3}$,
	Haibo Mi$^{2,3}$,
	Huaimin Wang$^{2,3}$, 
	Jian Zhu$^{4}$
}

\address{
	$^1$ Beijing University of Posts and Telecommunications, Automation Dept., Beijing, 100876, China\\        
	$^2$ National Key Lab of Parallel and Distributed Processing, Changsha, China\\
	$^3$ National University of Defense Technology, Changsha, China\\
	$^4$ Department of Linguistics, University of Michigan, Ann Arbor, USA\\
	kelele.xu@gmail.com\\
}
\begin{document}
%
\maketitle
\begin{abstract}
B-mode ultrasound tongue imaging is widely used in the speech production field. However, efficient interpretation is in a great need for the tongue image sequences. Inspired by the recent success of unsupervised deep learning approach, we explore unsupervised convolutional network architecture for the feature extraction in the ultrasound tongue image, which can be helpful for the clinical linguist and phonetics. By quantitative comparison between different unsupervised feature extraction approaches, the denoising convolutional autoencoder (DCAE)-based method outperforms the other feature extraction methods on the reconstruction task and the 2010 silent speech interface challenge. A Word Error Rate of 6.17\% is obtained with DCAE, compared to the state-of-the-art value of 6.45\% using Discrete cosine transform as the feature extractor. Our codes are available at https://github.com/DeePBluE666/Source-code1.

\end{abstract}
\section{Introduction}
\label{sec:intro}

The notion of developing a ``silent speech interface (SSI)'' have attracted great attention in recent years. SSI is often envisioned as a system which will convert silent speech related movements captured by motion tracking instruments back into texts or speech signals. Such a system has potential applications in a variety of situations, such as in extreme noisy environments, or in situations where non-vocalized communication is preferred, or in cases of patients whose vocalization functions are impaired by laryngectomy \cite{ji2018updating}. In previous works, prototypes of SSI have been developed using different articulatory motion capture instruments, including eletromagnetic articulography (EMA)\cite{fagan2008development}, electromyography (EMG) \cite{schultz2010modeling}, ultrasound tongue imaging (UTI) \cite{ji2018updating,hueber2010development}, non-audible murmur microphone \cite{tran2010improvement} and vibration sensors attached to the head and neck \cite{patil2010physiological}.

In the current study, we focus on the ultrasound based SSI. Ultrasound tongue imaging is one of the widely used tools in speech production research and clinical diagnostics of speech pathologies \cite{stone2005guide}. Compared with other articulatory motion tracking devices, which often require extra efforts to attach sensors to speakers' major articulators (e.g., EMA, EMG), ultrasound machine is safe, non-invasive, portable and low-cost, making it a desirable option for SSI systems. Ultrasound imaging can capture continuous tongue movements with frame rates ranging from 30-100 Hz, therefore containing rich speech-related information \cite{xu2016robust}. Though vocal source and lip movements cannot be recorded by ultrasound imaging, in an ultrasound-based SSI, ultrasound recordings can also be augmented with lip movement videos captured by a video camera \cite{ji2018updating,hueber2010development}. The downside, however, is that ultrasound images are high-dimensional and sparse, often suffering from low signal-to-noise ratio and contamination of speckled noises \cite{xu2016comparative}. For example, the presence of high contrast edges may obscure the underlying tongue shape \cite{stone2005guide}. Besides, ultrasound images can be distorted by rotation or relative displacement due to accidental head movements during the recording. Extracting invariant speech features from degraded ultrasound images therefore remains a key issue in the development of an ultrasound based SSI.

In previous studies, methods have been developed for extracting tongue shape features from ultrasound sequences, including both supervised and unsupervised methods. Different feature extraction methods are exploited extensively in works on ultrasound-based SSIs with impressive results on visual speech recognition, notably the Principal Component Analysis (PCA) based EigenTongue \cite{hueber2007eigentongue}, Discrete Cosine Transform (DCT) \cite{cai2011recognition},
and Deep Auto-encoder (DAE) \cite{ji2018updating}. In EigenTongue analysis, ultrasound image features are represented by the projection of ultrasound images onto basis vectors (EigenTongues), which are extracted from a set of typical images using PCA \cite{hueber2010development}. Similar to PCA, DCT also decomposes each ultrasound image into a linear combination of basis vectors but without the training step in PCA. It has been shown that the low frequency components of ultrasound images obtained with DCT are more effective than EigenTongue features \cite{cai2011recognition}. Recently, neural network based methods for feature extraction are also being developed and began to produce competitive results \cite{xu2017convolutional,jaumard2016articulatory}. In a study on multi-modal silent speech recognition, the auto-encoder based features have been shown to be comparable to DCT-based features but the auto-encoder based features can achieve similar performance with fewer feature vectors \cite{ji2018updating}.

In this paper, we explore potential of the denoising convolutional auto-encoder (DCAE) as an unsupervised feature extraction method for ultrasound image processing. To be more specific, DCAE is a variant of the supervised convolutional neural networks (CNN). CNNs, however, are trained only to learn filters able to extract features that can be used to reconstruct the input. Compared with the classic deep auto-encoder (DAE), DCAE may be better suited for processing ultrasound image because they fully utilize the properties of convolutional architecture, which have been proven to provide better results on noisy, shifted (translated) and corrupted image data.  Moreover, it is well known that ultrasound contains high level speckle noise, DCAE maybe better suitable to remove the noisy components in ultrasound images. Such convolutional auto-encoders can be trained in end-to-end manner with the unlabeled data.

The remainder of this paper is organized as follows: Section 2 describes the methodology for the unsupervised feature extraction in the ultrasound tongue image sequences. Section 3 presents the results on the reconstruction task and 2010 Silent Speech Interface Challenge, and Section 4 draw a conclusion in the end.

\section{Overview of auto-encoders}
\label{sec:cae}
In \cite{ji2018updating}, a deep auto-encoder based method of feature extraction, or dimension reduction, is proposed to process ultrasound tongue images. Indeed, auto-encoders provide a powerful framework for learning compact representations by encoding all of the information needed to reconstruct an image (here, which is a ultrasound tongue image frame) in a latent space. However, since the revolution of neural network, it is found that convolutional architectures perform better on image-related tasks, such as image classification \cite{krizhevsky2012imagenet} and medical image segmentation \cite{litjens2017survey}. Thus, we explore the potential of convolutional auto-encoders in the context of feature extraction from ultrasound tongue images.

\subsection{Deep Auto-encoder}
A deep auto-encoder(AE) is a variant of feed-forward neural network that learns to compress the input data into more compact representations\cite{liou2008modeling}, and then to reconstruct as closely as possible the original input data based on the learned compact representations. It is an unsupervised learning algorithm, which aims at setting the target values to be equal to the inputs. In this way, auto-encoder seeks to learn an approximation function $h_{W,b}(x)\approx x$, where $W$ and $b$ are the parameters. An auto-encoder consists of two symmetric but separable parts, the encoder and the decoder. And the training process is to learn the parameters that minimize the differences between original data and the reconstructed outputs through back-propagation.

\subsection{Denoising Auto-encoder}
\begin{figure}[ht]
	\centering
	\includegraphics[scale=0.3]{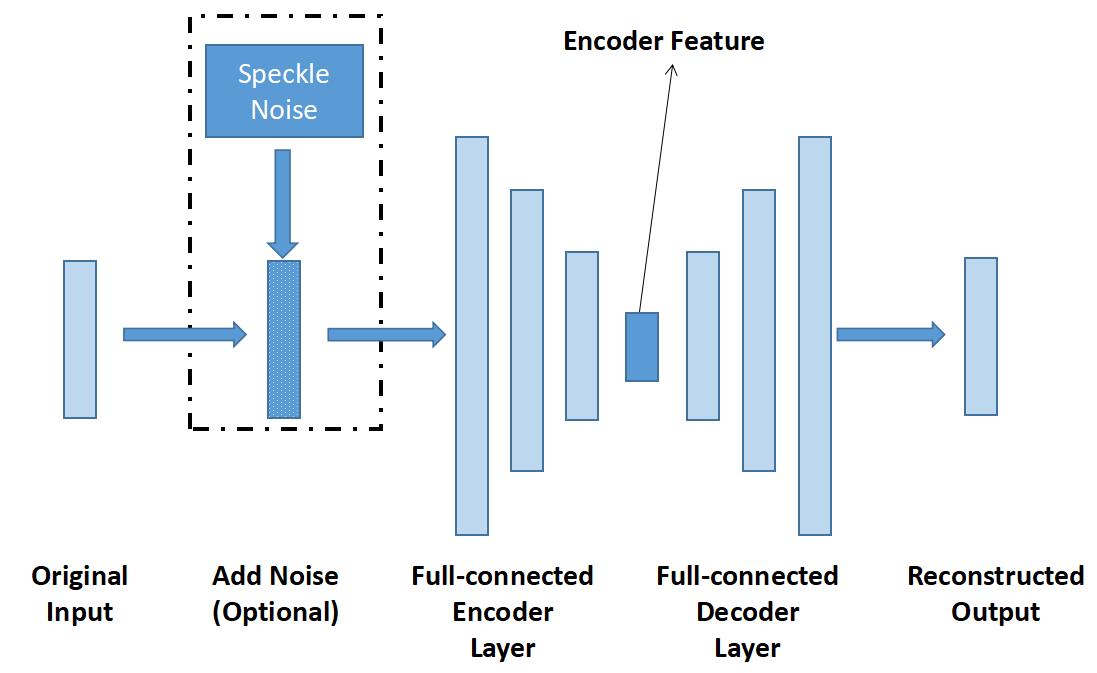}
	\caption{The structure of AE and DAE}
	\label{The structure of AE and DAE}
\end{figure}

As an extension of AE, denoising auto-encoders(DAE) learns to approximate the original input by training on the input vectors with noises. Denoising auto-encoder is designed to reconstruct the original data from the corrupted version of the original images \cite{vincent2010stacked}, the process of which forces the hidden layer to discover more robust features and prevents overfitting noises. In general, denoising auto-encoder aims to achieve two complementary goals: extracting most representative features from inputs, which are the ultrasound tongue images in this study, and minimizing the effect of noise applied to the input. However, the loss function only compares the output values with the original input values with noises. Fig\ref{The structure of AE and DAE}shows the structure of deep auto-encoder and denoising auto-encoder.

\subsection{Convolutional Auto-encoder}
Convolutional autoencoders (CAE) share the same overall architecture as classic AE with encoding and decoding layers but are tailored to processing images \cite{masci2011stacked}. Instead of having fully connected layers, the encoding layers consist of convolutional and max-pooling layers that project high dimensional input vectors into lower dimensional compact feature vectors, while the decoding layers are formed by convolutional layers as well as upsampling layers that upsample the feature vectors back to original size. As the convolution operation preserves the inner structure of image through local weight sharing, CAE is argued to be suitable for image denoising and image feature extraction \cite{gondara2016medical}.

\subsection{Denoising Convolutional Auto-encoder}
\begin{figure}[ht]
	\centering
	\includegraphics[scale=0.3]{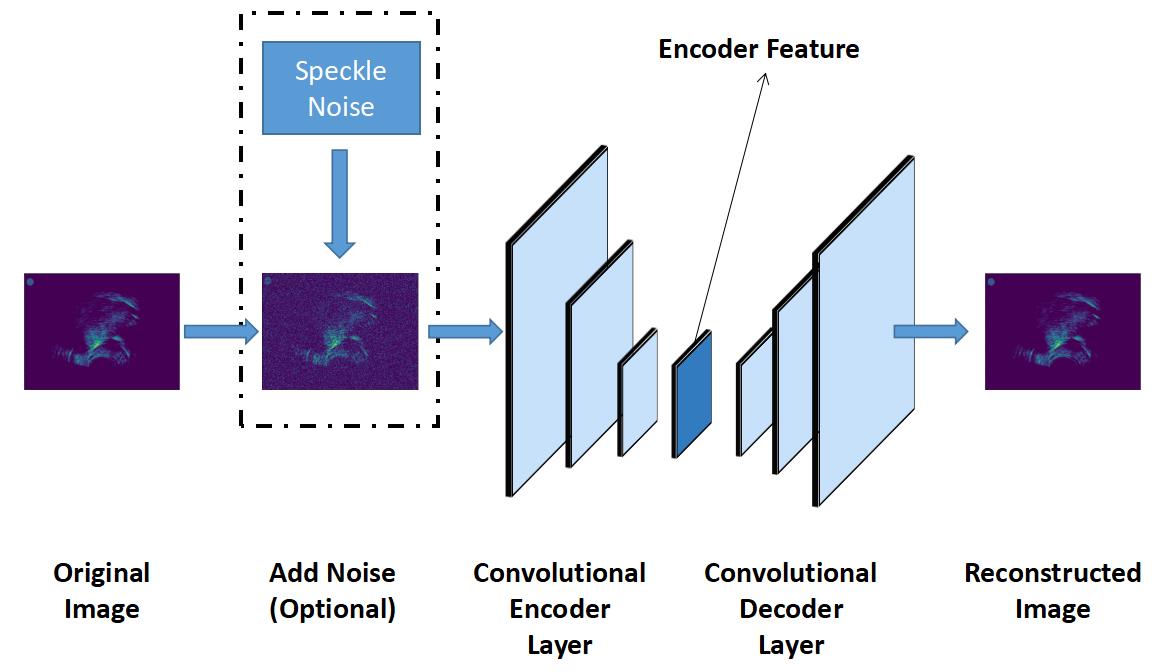}
	\caption{The structure of CAE and DCAE}
	\label{The structure of CAE and DCAE}
\end{figure}

Denoising Convolutional Auto-encoder(DCAE) is the improved version of CAE with the same convolutional architecture. Compared with CAE, DCAE can be more robust to corrupted inputs (such the ultrasound tongue images with high-level speckle noise), as the random noises are added to simulate noises in the ultrasound frames. An example structure of CAE and DCAE are shown in Fig.\ref{The structure of CAE and DCAE}.

\section{Quantitative comparison}
\subsection{Dataset and Experiment setup}
We evaluate different feature extraction methods using the 2010 Silent Speech Challenge dataset, which were recorded mid-sagittal ultrasound tongue images at a rate of 60 frames per second, using an acquisition light helmet that stabilizes a 4-8 MHz, 128-element, microconvex ultrasound probe beneath the speaker’s chin. The size of the raw images is 320$\times$240. The ROIs were selected, and all the ROIs were resized to 50$\times$60 pixels before the feature extraction using bi-cubic interpolation, with the goal to make results comparable with those in \cite{ji2018updating}.

In \cite{ji2018updating}, the deep auto-encoder architecture consists of 3 encoding layers and 3 decoding layers, in which each layer has 1000, 500, 250 hidden units. In this paper, deep AE and DCT are used as the benchmark of our experiments. As for CAE and DCAE, all encoder layers have 3 convolutional layers and 3 max-pooling layers. After the test using different architectures, we employ following architectures for CAE and DCAE, in which the size of first convolutional layer is $8\times8$ with 32 filters, the size of second convolutional layer is $6\times6$ with 16 filters, and the last convolutional layer has 2 filters with $4\times4$ convolution kernel. Depending on the dimensions of the encoder feature, we employed different sizes of max-pooling.

To train the DAEs and DCAEs, randomly generated speckle noises were added to the original input. Besides, we perform optimization with Adam \cite{kingma2014adam} using a fixed learning rate of 0.001 and fixed epoch number of 200.

\subsection{Reconstruction error comparison}
In this part, a quantitative comparison is carried out to assess the quality of ultrasound images reconstructed using deep auto-encoder, DCT, Denoising auto-encoder, CAE and Denoising CAE. We employ 50,000 training images and 30,000 images for test, all of which were randomly selected from the Silent Speech Challenge dataset.

Table \ref{table1} shows the reconstruction performances of different feature extraction methods. We report the mean square error (MSE)\cite{lehmann2006theory} and Complex Wavelet Structural Similarity Index (CW-SSIM)\cite{sampat2009complex} between original ultrasound tongue image and the reconstructed one.

\begin{table}[h]
	\centering 
	\caption{The performance of different Auto-encoder models with different structures and different encoder feature sizes.}
	\setlength{\tabcolsep}{0.5mm}{
		\begin{tabular}{|c|c|c|} 
			\hline
			$\begin{array}{c}\rm Size\ of \\\rm Encoder\ Features
			\end{array}$&$\begin{array}{c}\rm test \\\rm MSE
			\end{array}$&$\begin{array}{c}\rm test \\\rm cw\_ssim
			\end{array}$\\\hline
			\multicolumn{3}{|c|}{\textbf{\textit{DCT}}}\\\hline
			$1\times(5, 6)$&86.27&0.6286\\\hline
			$1\times(10, 6)$&89.03&0.6136\\\hline
			
			\multicolumn{3}{|c|}{\textbf{\textit{Deep AE}}}\\\hline
			$1\times(30, 1)$&87.92&0.6275\\\hline
			$1\times(60, 1)$&79.11&0.6241\\\hline
			
			\multicolumn{3}{|c|}{\textbf{\textit{Denoising AE}}}\\\hline
			$1\times(30, 1)$&79.03&0.6346\\\hline
			$1\times(60, 1)$&79.15&0.6313\\\hline
			
			\multicolumn{3}{|c|}{\textbf{\textit{CAE}}}\\\hline
			$1\times(5, 6)$&77.88&0.6435\\\hline
			$2\times(5, 6)$&71.76&0.6556\\\hline
			
			\multicolumn{3}{|c|}{\textbf{\textit{Denoising CAE}}}\\\hline
			$1\times(5, 6)$&77.21&0.6421\\\hline
			$2\times(5, 6)$&\textbf{71.28}&\textbf{0.6577}\\\hline
			
		\end{tabular}
		\label{table1}}
\end{table}

In the table, the first column shows the size of the encoder features. For example, $2\times(5, 6)$ denotes that: the encoder features (or feature maps) have 2 channels in the coder layer, and the shape of each feature map is $5\times6$. Thus, the length of the encoder features is 60. As can be seen from the table: (1) both CAE and DCAE models exhibit better reconstruction performance compared to deep AE models using the MSE and CW-SSIM metrics. The results may suggest that CAE can capture more representative pattern in the ultrasound tongue image. (2) Denoising AE provided better reconstruction performance with comparison to the traditional deep auto-encoder, and DCAE performed better than CAE in the same task. (3) DCT can provide similar reconstruction performance as the Deep auto-encoder with the length 30. (4) With the increase of the feature length, the reconstruction error can be reduced, except with DCT.

\begin{figure}[ht]
	\centering
	\includegraphics[scale=0.32]{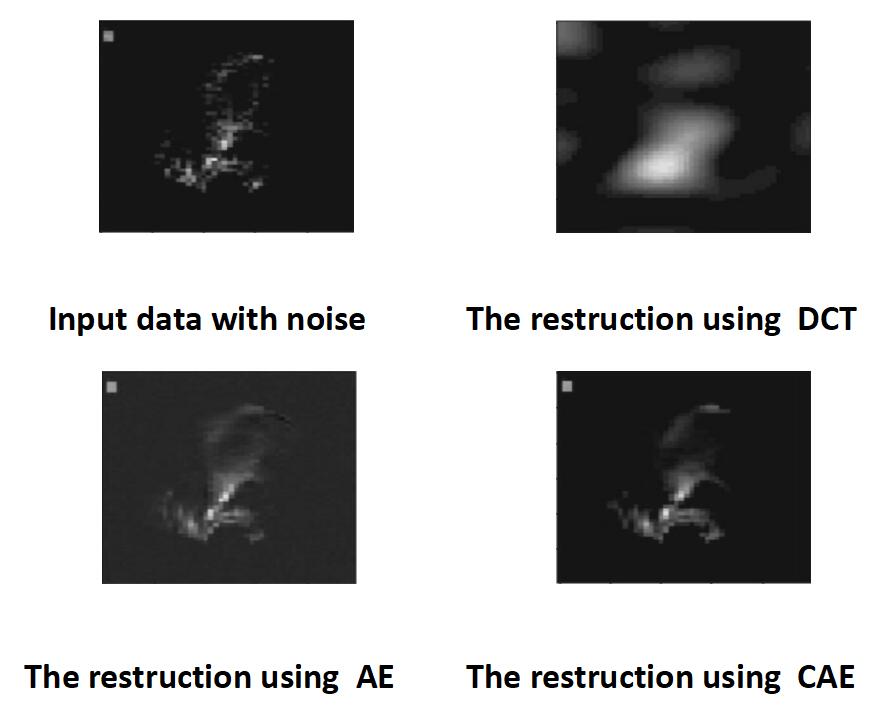}
	\caption{The reconstruction images using noisy inputs with different models. The length of each feature vector is 60.}
	\label{restruction_all}
\end{figure}

Moreover, reconstructed ultrasound images based on different features are generated for visual comparison, as shown in Fig \ref{restruction_all}.

\subsection{Silent speech challenge}
We also conducted a series of speech recognition experiments on the 2010 Silent Speech Challenge. We followed the same setting for the whole recognition system as in \cite{ji2018updating}, except with different feature extraction methods. As DNN-HMM outperforms the previous GMM-HMM, we only used the DMM-HMM model with the concatenated tongue and lip feature vectors of dimension K = 60 and 30, extracted using DCT, Deep AE, Denoising AE, CAEs, CAE and denoising CAE. We employed Kaldi for the implementation of the system, and Wall Street Journal (WSJ) language model was deployed in our recognition method.

Table 2 displays the outcomes of the DNN-HMM silent speech recognition system for different input features of either 30 or 60-dimensions resulting from DCT, CAE, and Denoising CAE respectively. The word error rate (WER) used to evaluate the recognition performance. According to Table \ref{SSC}, WER is lower for both DCT and CAE based features than for Deep AE and Denoising AE based features, but the lowest WER is achieved with DCAE based features. These results suggest that CAEs encode more speech-related information than AEs.

\begin{table}[h]
	\centering 
	\caption{The Word Error Rate of the Silent Speech Challenge with different feature extractors.}
	\begin{tabular}{|c|c|c|}
		\hline
		Length & 60 & 30\\\hline
		DCT & 6.58 & 6.45\\\hline
		Deep AE& 7.37& 7.72\\\hline
		Denoising AE& 7.26 & 7.54\\\hline
		CAE& 6.22 & 6.34\\\hline
		Denoising CAE& \textbf{6.17}& \textbf{6.24}\\\hline
	\end{tabular}
	\label{SSC}
\end{table}

\section{CONCLUSION}
\label{sec:conclusion}
In this study, we introduce a new feature extraction method, Denoising CAE, for extracting speech-related features from ultrasound images, in which the convolutional layers should make the method robust against the distortion resulting from head movements, variability due to different imaging systems and speckle noises inherent in the ultrasound images. We believe this is the first attempt to use denoising CAE for feature extraction in the ultrasound tongue image. Four methods for feature extraction from ultrasound images have been presented and compared. The visual-articulatory modeling with denoising CAE gives better results than those obtained using non-convolutional architecture-based auto-encoder as input. The denoising CAE method better preserves the spatial information in ultrasound images, and is less prone to failures due to different ultrasound instruments, thereby allowing the dynamic nature of speech to be taken into account in a natural way. The sequence auto-encoder and the use of alternative dynamic process modeling techniques (Recurrent Neural Networks) are currently underway.

\section{ACKNOWLEDGMENT}
This work was supported by the National Grand R\&D Plan(Grant No. 2016YFB1000101). The authors would like to acknowledge useful discussions and helpful suggestions from Zhifeng Gao (Microsoft Asia). The authors also thank for Chengrui Wu for providing the code for the 2010 silent speech challenge.

\bibliographystyle{IEEEbib}
\bibliography{ref}

\begin{thebibliography}{10}

\bibitem{ji2018updating}
Yan Ji, Licheng Liu, Hongcui Wang, Zhilei Liu, Zhibin Niu, and Bruce Denby,
\newblock ``Updating the silent speech challenge benchmark with deep
  learning,''
\newblock {\em Speech Communication}, vol. 98, pp. 42--50, 2018.

\bibitem{fagan2008development}
Michael~J Fagan, Stephen~R Ell, James~M Gilbert, E~Sarrazin, and Peter~M
  Chapman,
\newblock ``Development of a (silent) speech recognition system for patients
  following laryngectomy,''
\newblock {\em Medical engineering \& physics}, vol. 30, no. 4, pp. 419--425,
  2008.

\bibitem{schultz2010modeling}
Tanja Schultz and Michael Wand,
\newblock ``Modeling coarticulation in emg-based continuous speech
  recognition,''
\newblock {\em Speech Communication}, vol. 52, no. 4, pp. 341--353, 2010.

\bibitem{hueber2010development}
Thomas Hueber, Elie-Laurent Benaroya, G{\'e}rard Chollet, Bruce Denby,
  G{\'e}rard Dreyfus, and Maureen Stone,
\newblock ``Development of a silent speech interface driven by ultrasound and
  optical images of the tongue and lips,''
\newblock {\em Speech Communication}, vol. 52, no. 4, pp. 288--300, 2010.

\bibitem{tran2010improvement}
Viet-Anh Tran, G{\'e}rard Bailly, H{\'e}l{\`e}ne L{\oe}venbruck, and Tomoki
  Toda,
\newblock ``Improvement to a nam-captured whisper-to-speech system,''
\newblock {\em Speech communication}, vol. 52, no. 4, pp. 314--326, 2010.

\bibitem{patil2010physiological}
Sanjay~A Patil and John~HL Hansen,
\newblock ``The physiological microphone (pmic): A competitive alternative for
  speaker assessment in stress detection and speaker verification,''
\newblock {\em Speech Communication}, vol. 52, no. 4, pp. 327--340, 2010.

\bibitem{stone2005guide}
Maureen Stone,
\newblock ``A guide to analysing tongue motion from ultrasound images,''
\newblock {\em Clinical linguistics \& phonetics}, vol. 19, no. 6-7, pp.
  455--501, 2005.

\bibitem{xu2016robust}
Kele Xu, Yin Yang, Maureen Stone, Aurore Jaumard-Hakoun, Cl{\'e}mence
  Leboullenger, G{\'e}rard Dreyfus, Pierre Roussel, and Bruce Denby,
\newblock ``Robust contour tracking in ultrasound tongue image sequences,''
\newblock {\em Clinical linguistics \& phonetics}, vol. 30, no. 3-5, pp.
  313--327, 2016.

\bibitem{xu2016comparative}
Kele Xu, Tam{\'a}s G{\'a}bor~Csap{\'o}, Pierre Roussel, and Bruce Denby,
\newblock ``A comparative study on the contour tracking algorithms in
  ultrasound tongue images with automatic re-initialization,''
\newblock {\em The Journal of the Acoustical Society of America}, vol. 139, no.
  5, pp. EL154--EL160, 2016.

\bibitem{hueber2007eigentongue}
Thomas Hueber, Guido Aversano, G{\'e}rard Chollet, Bruce Denby, G{\'e}rard
  Dreyfus, Yacine Oussar, Pierre Roussel-Ragot, and Maureen Stone,
\newblock ``Eigentongue feature extraction for an ultrasound-based silent
  speech interface.,''
\newblock in {\em ICASSP (1)}, 2007, pp. 1245--1248.

\bibitem{cai2011recognition}
Jun Cai, Bruce Denby, Pierre Roussel, G{\'e}rard Dreyfus, and Lise
  Crevier-Buchman,
\newblock ``Recognition and real time performances of a lightweight ultrasound
  based silent speech interface employing a language model,''
\newblock in {\em Twelfth Annual Conference of the International Speech
  Communication Association}, 2011.

\bibitem{xu2017convolutional}
Kele Xu, Pierre Roussel, Tam{\'a}s~G{\'a}bor Csap{\'o}, and Bruce Denby,
\newblock ``Convolutional neural network-based automatic classification of
  midsagittal tongue gestural targets using b-mode ultrasound images,''
\newblock {\em The Journal of the Acoustical Society of America}, vol. 141, no.
  6, pp. EL531--EL537, 2017.

\bibitem{jaumard2016articulatory}
Aurore Jaumard-Hakoun, Kele Xu, Cl{\'e}mence Leboullenger, Pierre
  Roussel-Ragot, and Bruce Denby,
\newblock ``An articulatory-based singing voice synthesis using tongue and lips
  imaging,''
\newblock in {\em ISCA Interspeech 2016}, 2016, vol. 2016, pp. 1467--1471.

\bibitem{krizhevsky2012imagenet}
Alex Krizhevsky, Ilya Sutskever, and Geoffrey~E Hinton,
\newblock ``Imagenet classification with deep convolutional neural networks,''
\newblock in {\em Advances in neural information processing systems}, 2012, pp.
  1097--1105.

\bibitem{litjens2017survey}
Geert Litjens, Thijs Kooi, Babak~Ehteshami Bejnordi, Arnaud Arindra~Adiyoso
  Setio, Francesco Ciompi, Mohsen Ghafoorian, Jeroen~AWM van~der Laak, Bram
  Van~Ginneken, and Clara~I S{\'a}nchez,
\newblock ``A survey on deep learning in medical image analysis,''
\newblock {\em Medical image analysis}, vol. 42, pp. 60--88, 2017.

\bibitem{liou2008modeling}
Cheng-Yuan Liou, Jau-Chi Huang, and Wen-Chie Yang,
\newblock ``Modeling word perception using the elman network,''
\newblock {\em Neurocomputing}, vol. 71, no. 16-18, pp. 3150--3157, 2008.

\bibitem{vincent2010stacked}
Pascal Vincent, Hugo Larochelle, Isabelle Lajoie, Yoshua Bengio, and
  Pierre-Antoine Manzagol,
\newblock ``Stacked denoising autoencoders: Learning useful representations in
  a deep network with a local denoising criterion,''
\newblock {\em Journal of machine learning research}, vol. 11, no. Dec, pp.
  3371--3408, 2010.

\bibitem{masci2011stacked}
Jonathan Masci, Ueli Meier, Dan Cire{\c{s}}an, and J{\"u}rgen Schmidhuber,
\newblock ``Stacked convolutional auto-encoders for hierarchical feature
  extraction,''
\newblock in {\em International Conference on Artificial Neural Networks}.
  Springer, 2011, pp. 52--59.

\bibitem{gondara2016medical}
Lovedeep Gondara,
\newblock ``Medical image denoising using convolutional denoising
  autoencoders,''
\newblock in {\em Data Mining Workshops (ICDMW), 2016 IEEE 16th International
  Conference on}. IEEE, 2016, pp. 241--246.

\bibitem{kingma2014adam}
Diederik~P Kingma and Jimmy Ba,
\newblock ``Adam: A method for stochastic optimization,''
\newblock {\em arXiv preprint arXiv:1412.6980}, 2014.

\bibitem{lehmann2006theory}
Erich~L Lehmann and George Casella,
\newblock {\em Theory of point estimation},
\newblock Springer Science \&amp; Business Media, 2006.

\bibitem{sampat2009complex}
Mehul~P Sampat, Zhou Wang, Shalini Gupta, Alan~Conrad Bovik, and Mia~K Markey,
\newblock ``Complex wavelet structural similarity: A new image similarity
  index,''
\newblock {\em IEEE transactions on image processing}, vol. 18, no. 11, pp.
  2385--2401, 2009.

\end{thebibliography}

\end{document}